 \RequirePackage[2020-02-02]{latexrelease}
\documentclass[floatfix,aps,prl,twocolumn,groupedaddress,showpacs,english]{revtex4}
\usepackage{color,soul}
\usepackage{placeins}
\usepackage[T1]{fontenc}
\usepackage[latin9]{inputenc}
\usepackage{amsmath}
\usepackage{amssymb}
\usepackage{graphicx}
\usepackage{subfig}
\usepackage{xcolor}

\usepackage{caption}
\usepackage{appendix}
\usepackage{wasysym}
\usepackage[version=4]{mhchem}
\usepackage{collcell}

\makeatletter
\@ifundefined{showcaptionsetup}{}{%
 \PassOptionsToPackage{caption=false}{subfig}}
\usepackage{subfig}
\makeatother

\usepackage{babel}
\begin{document}
\title{%Passive State Transfer/ 
Efficient General Waveform Catching by a cavity   at a Virtual Absorbing
Exceptional Point}
\author{Asaf Farhi\textsuperscript{1}, Wei Dai\textsuperscript{1}, Seunghwi Kim\textsuperscript{2}, Andrea Alu\textsuperscript{2,3}, and Douglas Stone\textsuperscript{1,4}}
%\date{\today}
\affiliation{\textsuperscript{1}Department of Applied Physics, Yale University, New Haven, Connecticut
06520, USA}\vspace*{2cm}
\affiliation{\textsuperscript{2} Photonics Initiative, Advanced Science Research Center, City University
of New York, New York, New York 10031, USA}
\affiliation{\textsuperscript{3}Physics Program, Graduate Center, City University of New York, New
York, New York 10016, USA}
\affiliation{\textsuperscript{4}Yale Quantum Institute, Yale University, New Haven, Connecticut 06520,
USA}

\begin{abstract}
State transfer and photon detection are fundamental processes that have direct
implications in fields such as quantum computing and photonic circuits. However, while naturally emitted photons decay exponentially in time, to perfectly capture a photon its envelope should increase exponentially to match the time-reversed response of the absorbing cavity. 
%In order to achieve high efficiency in quantum state transfer the
%wavepacket of the photon is usually modulated to enhance the photon catching. 
Here we show that a cavity at a virtual absorbing
exceptional point captures additional temporal orders of
an incoming waveform, resulting in efficient passive state transfer and photon
detection. 
This approach paves the way for state transfer
 at optical frequencies and efficient detection of a spontaneously emitted photon.

% This approach holds potential to allow quantum state transfer
% to operate at optical frequencies and .

\end{abstract}
\maketitle

Photon transfer between cavities is a fundamental process in both classical and quantum networks at microwave and optical frequencies.
Quantum networks composed of spatially separated nodes that interact via flying photons \cite{kimbleQuantumInternet2008,northupQuantumInformationTransfer2014} are a topic of long-standing interest for applications such as distributed quantum computation \cite{monroeLargescaleModularQuantumcomputer2014,jiangDistributedQuantumComputation2007} and quantum communication \cite{duanLongdistanceQuantumCommunication2001,childressFaultTolerantQuantumCommunication2006}. 
%In the past two decades, a number of experiments demonstrated remote entanglement or teleportation in quantum networks, mostly based on . In contrast, 
Deterministic protocols based on direct state transfer \cite{ciracQuantumStateTransfer1997} were shown to achieve larger entanglement rates compared with probabilistic protocols \cite{chouMeasurementinducedEntanglementExcitation2005,moehringEntanglementSingleatomQuantum2007,hofmannHeraldedEntanglementWidely2012,bernienHeraldedEntanglementSolidstate2013,olmschenkQuantumTeleportationDistant2009}, yet requiring strong light-matter interactions for transfer efficiency \cite{ritterElementaryQuantumNetwork2012}. In recent years, direct state transfer has  been realized in superconducting circuit systems \cite{kurpiersDeterministicQuantumState2018,axlineOndemandQuantumState2018,campagne-ibarcqDeterministicRemoteEntanglement2018}, featuring on-demand "pitch-and-catch" of propagating microwave quantum signal with high efficiency. 
In such experiments a resonant cavity coupled to an atom (or artificial atom) is commonly utilized at a quantum node, in both microwave \cite{flurinSuperconductingQuantumNode2015,yinCatchReleaseMicrowave2013} and optical \cite{ritterElementaryQuantumNetwork2012,reisererCavitybasedQuantumNetworks2015} networks. Capturing a propagating wave packet by a cavity is normally the first step in flying photon processing.

Photon detection is another topic of paramount importance in quantum information applications as well as for probing processes such as atomic or molecular spontaneous emission, electron spin resonance, nuclear magnetic resonance, and fluorescence \cite{goy1983observation,albertinaleDetectingSpinsTheir2021,wangSingleelectronSpinResonance2023,graham2022multi}.  In particular, detection of atomic fluorescence, in which the spontaneous emission frequency equals the excitation frequency, has become an integral part of quantum computation with neutral atoms \cite{graham2022multi,browaeys2020many}. A key requirement of photon detectors is large detection efficiency of an incoming photon \cite{hadfield2009single}. One of the common implementations of a photon detector also relies on resonant cavities \cite{albertinaleDetectingSpinsTheir2021,wangSingleelectronSpinResonance2023}.

The most efficient way to load a resonator is with an input wave at its complex absorbing eigenfrequency, which has an \emph{increasing} exponential waveform, see Fig. 1 (b), as demonstrated in several experiments at the microwave \cite{wennerCatchingTimeReversedMicrowave2014,palomakiCoherentStateTransfer2013,flurinSuperconductingQuantumNode2015,linDeterministicLoadingMicrowaves2022} and optical frequencies \cite{baderEfficientCouplingOptical2013,liuEfficientlyLoadingSingle2014,baranov2017coherent}. This scatterless excitation of lossless cavities is the complex frequency generalization of coherent perfect absorber
(CPA), the time reversal of a laser \cite{chong2010coherent,noh2012perfect}.
However, for state transfer this needs to involve also the pitch process, which makes it experimentally challenging \cite{khanahmadi2023multimode}, and photon waveforms emitted in natural processes usually do not have this waveform. A common approach for efficient state transfer in cavity Quantum Electrodynamics (cQED) experiments
\cite{korotkovFlyingMicrowaveQubits2011,seteRobustQuantumState2015} is to generate and capture a time-symmetrically-shaped wave packet \cite{kurpiersDeterministicQuantumState2018,axlineOndemandQuantumState2018,campagne-ibarcqDeterministicRemoteEntanglement2018} by dynamically modulating both the emitter and receiver nodes with fast flux tuning \cite{srinivasanTimereversalSymmetrizationSpontaneous2014,yinCatchReleaseMicrowave2013,pierreStorageOndemandRelease2014}, microwave-controlled parametric conversion \cite{pfaffControlledReleaseMultiphoton2017,flurinSuperconductingQuantumNode2015}, or stimulated Raman process \cite{pechalMicrowaveControlledGenerationShaped2014,kurpiersDeterministicQuantumState2018,campagne-ibarcqDeterministicRemoteEntanglement2018}. Another interesting direction of temporal switching has allowed unitary excitation transfer between coupled cavities \cite{mazor2021unitary}. 
% AF no space
Despite these techniques of active pulse shaping of flying photons, it remains important to passively catch photons that are naturally emitted from cavities or in  spontaneous emission \cite{houckGeneratingSingleMicrowave2007,kindelGenerationEfficientMeasurement2016,heugel2010analogy}. Such emitted photons have a \emph{decaying} exponential waveform, see Fig. 1 (a), and when passively caught by a standard receiving cavity, the efficiency is rather poor, of typically  only 60\%   \cite{wennerCatchingTimeReversedMicrowave2014}. Clearly, efficient passive capture of naturally emitted photons holds the potential to allow quantum state transfer to operate at optical frequencies and exceptional detection of processes such as spontaneous emission. 

\begin{figure}
\includegraphics[width=8cm]{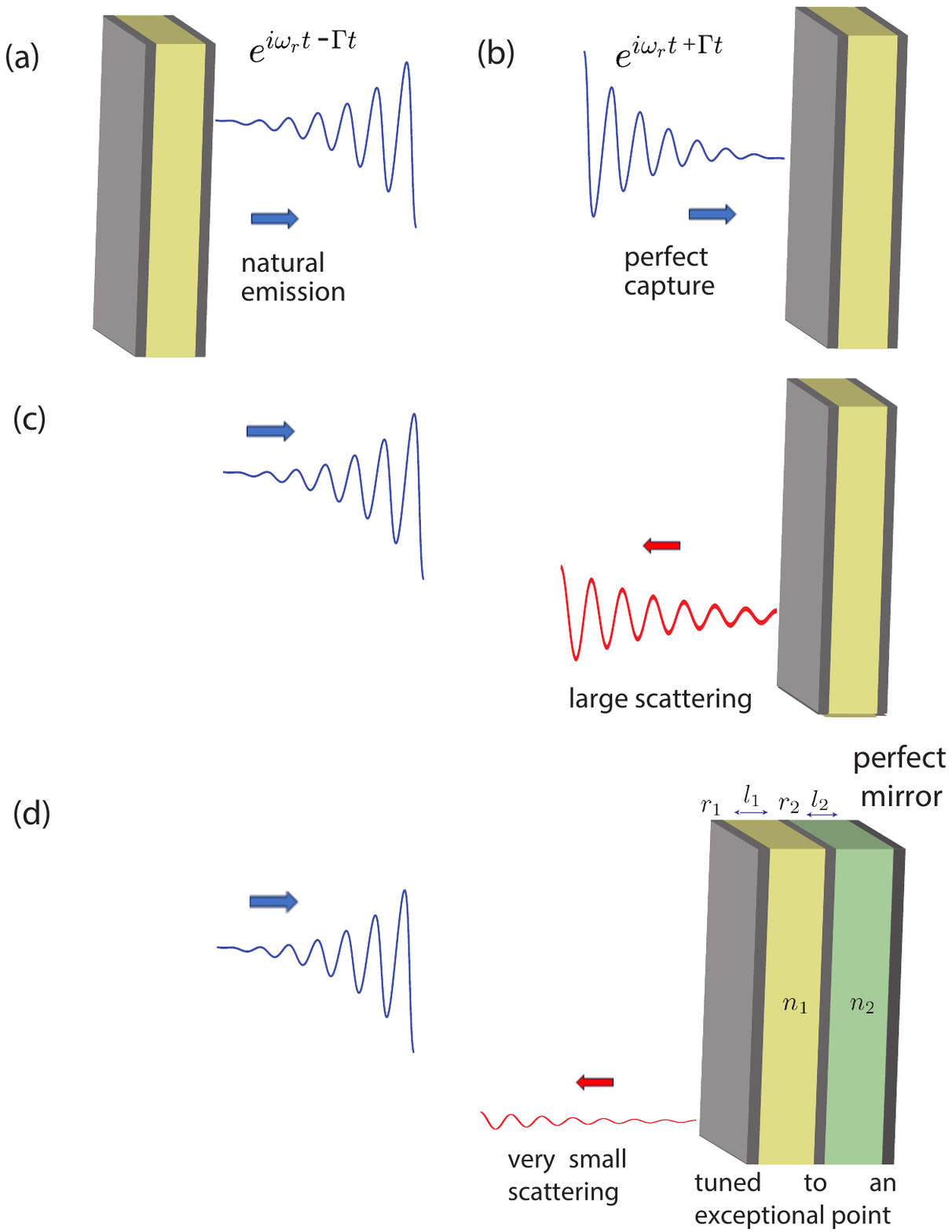}

\caption{(a) Naturally emitted waveform from a cavity that decays (increases) exponentially in time (space). A waveform with the same temporal dependency is emitted in spontaneous emission by an atom or molecule. (b) The perfectly captured waveform that increases (decays) exponentially in time (space). (c) Passive state transfer with standard cavities. Since the naturally emitted and perfectly captured waves do not match, 40\% of the incident field is scattered. (d) Passive state transfer with a cavity system tuned to a virtual absorbing exceptional point. Since the EP cavity captures another order in time of the incoming waveform, there is very small scattering.}
\end{figure}

%The absorption real-$\omega$ spectrum of a cavity can be altered by designing it to operate at an exceptional point (EP)  \cite{bender1998real,makris2008beam,moiseyev2011non,miri2019exceptional,ruter2010observation} of CPAs, in which CPA eigenvalues and eigenmodes coalesce \cite{sweeney2019perfectly,wang2021coherent}.

% Transition is weak. The previous paragraph says that a cavity cannot entirely capture naturally emitted photons. And suddenly, you start with the quartic response of CPA EP. But I would talk about how CPA EPs can help the absorption of naturally emitted photons first -- like its capability for absorbing high-order waveforms, so some arbitrary waveforms can be caught, indicating the naturally emitted photons are. At this moment, my feeling is this first sentence begins without justifying the agenda.

% AF you are right but we need to give credit to that work and Doug also edited the paragraph in this direction and we need to converge. 

It was recently shown that
by designing a cavity to operate at an exceptional point (EP)  \cite{bender1998real,makris2008beam,moiseyev2011non,miri2019exceptional,ruter2010observation} of CPAs, in which CPA eigenvalues and eigenmodes coalesce \cite{sweeney2019perfectly,wang2021coherent},  the
absorption spectrum on the real-$\omega$ axis becomes quartic \cite{sweeney2019perfectly,wang2021coherent}.
More recently, the time-domain properties of CPA EPs were  studied both for real and complex $\omega$ (virtual CPA EP) \cite{farhi2022excitation} and it was shown that they extend the class of waveforms, which can be perfectly absorbed, namely $E\propto\left(vt-z\right)^{m}e^{i\left(kz-\omega t\right)},$
where $m$ is the EP order \cite{farhi2022excitation}. It was demonstrated that these high-order waveforms have dramatically improved performance
in wave capturing  \cite{farhi2022excitation}, with an experimental realization in electric circuits \cite{mekawy2023observation}. Interestingly, all the waveforms within this class have
a frequency content at a single $\omega$ for infinite time pulses i.e., at $z=0$  $\mathcal{F}\left(t^{m}e^{i\omega_1t}\right)=\delta^{\left(m\right)}\left(\omega-\omega_{1}\right).$

%Exciting a cavity with an exponentially increasing wave was demonstrated experimentally in the quantum regime and resulted in a high efficiency that is above the threshold  required for some of the quantum computing applications \cite{wennerCatchingTimeReversedMicrowave2014}. However, the naturally emitted wave from a pitching cavity decays exponentially in time and can be captured by another cavity with only 55\% efficiency \cite{wennerCatchingTimeReversedMicrowave2014}. To improve the efficiency a variety of protocols to modulate the emitted wave have been suggested and implemented experimentally \cite{campagne-ibarcqDeterministicRemoteEntanglement2018}. 

\begin{figure}
\includegraphics[width=8cm]{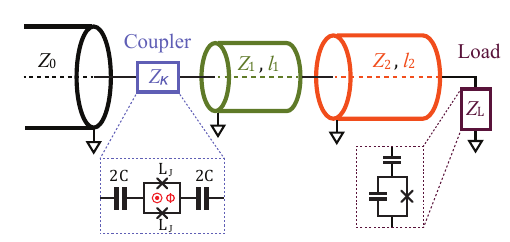}
\caption{A schematic of the setup: a flux-controlled coupler and two transmission lines (microwave analogue of the optical cavities) loaded by a transmon, tuned to an absorbing virtual exceptional point. The coupler is composed of two capacitors each of capacitance $2C$ and a SQUID, which can be modeled by an inductance $L = L_{J}/{\cos{(\Phi/\phi_0)}}$. }
\end{figure}

Here we first show that a cavity at a CPA EP captures additional orders in time of \emph{any} waveform. We then present a semianalytical approach to tune a catching port that is of the type used in quantum information processing to a virtual CPA EP. We next show that when this system is at an EP, it significantly improves the efficiency in the catch process of the waveforms $e^{i(\omega t-kz)},\,\,(vt-z)e^{i(\omega t-kz)}$ and can address additional applications such as good logic gates and measurements that have been challenging \cite{wennerCatchingTimeReversedMicrowave2014}.  Finally, we demonstrate that by utilizing such a cavity, the pitch and catch process can be performed without modifying the naturally emitted wave i.e., passive state transfer, at high efficiency (see Fig. 1 d) and discuss the advantage for single photon detection.

% Goals:
% \begin{itemize}
% \item Improve the efficiency of the catch process with an absorbing virtual
% exceptional point to address additional applications such as good
% logic gates and measurements \cite{wenner2014catching}. 
% \item Pitch and catch process without modifying the signal: passive state
% transfer, which may enable applications at optical frequencies.
% \item Another application: photon detection, also emitted by atoms and molecules
% e.g., spontaneous emission \cite{heugel2010analogy}.
% \end{itemize}

%\subsection*{Transferring a quantum state without modifying the signal}
We consider a single-port setup composed of two transmission line sessions loaded by a capacitively-coupled transmon qubit, and a flux-controled coupler which is typical in cQED experiments, see Fig. 2. 
For room-temperature applications \cite{graham2022multi} the transmon could be replaced by a standard photon detector; when such a detector already has several layers it can also replace the transmission lines.
We will now show that a first-order real or complex-$\omega$ CPA EP enables the capture of any waveform up to linear order in time. Clearly, waveforms with a significant linear order
term will substantially benefit from such an EP. We focus on naturally emitted waveforms which are of practical importance, can have a significant linear term, and their passive catching is somewhat counterintuitive. We start by showing that a cavity at a CPA EP captures  additional orders in time of these waveforms.

%\comt{This paragraph is too colloquial, need polishing}
% AF added a connecting sentence in the previous paragraph
Let us assume that we do not modulate the coupling in the pitch process.
Thus, the emitted photon has the form $e^{-\Gamma t+i\omega_{r}t}.$ % I think we should define \Gamma and \omega_r like, where $\Omega$ and $\omega_r$ are the resonant frequency and decay rate of the emitting cavity, respectively.
We consider a cavity
at a general complex-$\omega$ CPA EP in the catch process with the same $\omega_{r}$
as the emitting cavity but with $\Gamma_2.$ At  
a CPA EP any signal of
the form $e^{\Gamma_{2}t}\left(a+bt\right)$ can be perfectly captured (for any $a,b)$ \cite{farhi2022excitation}. We Taylor expand the naturally-emitted wave and the above-mentioned form and obtain that there exist $a,b$ for which they are equal to linear order
\begin{equation}e^{-\Gamma t}\approx\ensuremath{e^{\Gamma_{2}t}(1-(\Gamma+\Gamma_{2})t)},\end{equation} which means that  the naturally-emitted wave will be perfectly captured to linear order.  Similarly, at an exceptional point with three coalescing eigenmodes and eigenfrequencies (EP3), the naturally-emitted wave will be perfectly captured up to quadratic order, which implies longer  times
\begin{equation}
e^{-\Gamma t}\approx e^{\Gamma_{2}t}\left[1-(\Gamma+\Gamma_{2})t+\frac{1}{2}\left(\Gamma+\Gamma_{2}\right)^{2}t^{2}\right].
\end{equation}
Clearly, this can be generalized to any waveform and higher-order EPs. It is important to note that this is a \emph{transient} effect and that the polynomial $a+bt+..$ expands the function obtained by dividing the incoming wave by $e^{\Gamma_{2}t}.$
%, at a CPA EP  $e^{\Gamma_{2}t}$ and especially high-order waveforms are more efficiently captured compared with a CPA \cite{farhi2022excitation}. 
This greatly enhances the capturing efficiency, as we show below. 

We now present an approach to calculate EPs for our setup, which reduces a highly-complex calculation that is usually approximate to an eigenvalue equation with a very accurate solution.
We denote the  transmission line impedances by $Z_1,Z_2$ and the characteristic impedance by $Z_0$ and write the reflection coefficients at the $Z_{0}-Z_{1},Z_{1}-Z_{2},$
and $Z_{2}-Z_{3}$  as follows:
\[
r_{1}=\frac{Z_{k}\left(\omega\right)+Z_{1}-Z_{0}}{Z_{k}\left(\omega\right)+Z_{1}+Z_{0}},\,\,
r_{2}=\frac{Z_{2}-Z_{1}}{Z_{2}+Z_{1}},\,\,r_{3}=\frac{Z_{L}\left(\omega\right)-Z_{2}}{Z_{L}\left(\omega\right)+Z_{2}},
\]
where we include the coupler and load
with the impedances $Z_{k}\left(\omega\right)$ and $Z_{L}\left(\omega\right)$
at the $Z_{0}-Z_{1}$ and $Z_{2}-Z_{3}$ interfaces and the transmon load could be modeled by a lumped-element circuit.
By imposing boundary conditions, we obtain the total reflection coefficient
\[
r=\frac{r_{1}\left(\omega\right)\left(-r_{2}r_{3}e^{2ikl_{2}}+1\right)+e^{2ikl_{1}}\left(r_{2}-r_{3}\left(\omega\right)e^{2ikl_{2}}\right)}{1-r_{2}r_{3}e^{2ikl_{2}}+r_{1}\left(\omega\right)e^{2ikl_{1}}\left(r_{2}-r_{3}\left(\omega\right)e^{2ikl_{2}}\right)}.
\]
where $l_{1}$ and $l_{2}$ are the lengths of the first and second transmission
lines, respectively. Note that transmission lines can be designed with their impedances different while maintaining the same propagation speeds~\cite{pozar2011microwave}. For simplicity, we set the propagation speed to $c.$

It has been shown that the conditions for a CPA EP are  $r=0$ and $\frac{dr}{d\omega}=0$, which correspond to $g=0$ and $\frac{dg}{d\omega}=0$, where $g$ represents the numerator of $r$ ~\cite{farhi2022excitation}. 
Let us first set $g=0$ to obtain $r_{2}$ as follows:
\begin{gather}
%g=r_{1}\left(\omega\right)\left(-r_{2}r_{3}e^{2ikl_{2}}+1\right)+e^{2ikl_{1}}\left(r_{2}-r_{3}\left(\omega\right)e^{2ikl_{2}}\right)=0,\nonumber\\
r_{2}=\frac{r_{3}\left(\omega\right)e^{2ikl_{1}}e^{2ikl_{2}}-r_{1}\left(\omega\right)}{e^{2ikl_{1}}-r_{3}r_{1}\left(\omega\right)e^{2ikl_{2}}}.
\end{gather}
Since $r_{2}$ is real we express from
$\mathrm{Im}\left(r_{2}\right)=0,$
 e.g., $Z_{1}\left(\omega,r_{3}\left(\omega\right),l_{1},l_{2},Z_{k}\left(\omega\right)\right),$
which we substitute in $r_2$ and 
the equation formed by equating the two expressions for $r_2.$
From this equation we
find $Z_{2}\left(\omega,r_{3}\left(\omega\right),l_{1},l_{2},Z_{k}\left(\omega\right)\right)$
and then substitute $Z_{1},Z_{2},$ and $r_{2}$ in $dg/d\omega=0$
to get the EP equation
\begin{equation}
\frac{dg}{d\omega}\left(\omega,Z_{L}\left(\omega\right),l_{1},l_{2},Z_{k}\left(\omega\right)\right)=0.
\end{equation}
%We calculate the virtual CPA EP of the two transmission lines and coupler system. 
% Tbh, I'm not sure I fully understand the process in Eq. (4). So the reason that you did like that is to reduce the number of the parameters inside $\frac{dg}{d\omega} = 0$. g is a function of \omega, Z_k, Z_0,Z_1,Z_2, Z_L, l_1, and l_2. However, we can get r_2 in Eq. (3), and this r_2 is a function of Z_1 and Z_2. As the impedances of the transmission lines are real, Im(r_2) is 0. So we have three equations for r_2 -- 1) r_2 in Eq.(3), 2) r_2 = \frac{Z_2 - Z_1}{Z_2 + Z_1} and 3) Im(r_2) =0. Thus we can reduce the parameters by two in this case Z_1 and Z_2. Is it what you've done? Do you think you can simplify the description a bit? 

For concreteness, we assume a linear response, neglecting the nonlinearity from the weakly-coupled transmon and the coupler, which is justified for the case of a single photon. We also assume that the transmon is decoupled from the transmission lines during the drive \cite{wennerCatchingTimeReversedMicrowave2014}, which implies $r_3=-1.$ We model the coupler as a lumped $LC$ element (whose inductance is external-flux-dependent), with the impedance given by
$
Z_{k}=j\frac{\omega^{2}LC-1}{\omega C}.
$
While the practical schemes to implement the coupler are mostly at the microwave, recent works suggested an analogy between circuit elements and optical components \cite{engheta2005circuit, engheta2007circuits, alu2008tuning, schnell2009controlling, yao2013broad, liu2013individual, aouani2014third}. In addition, there was recent progress in optical switching, which can rapidly switch off the coupler \cite{venkataraman2011few,ren2011nanostructured}.  
%The reflection $r_{1}$ from the coupler and the $Z_{0}-Z_{1}$ interface reads 
%\[
%r_{1}=\frac{Z_{k}\left(\omega\right)+Z_{1}-Z_{0}}%{Z_{k}\left(\omega\right)+Z_{1}+Z_{0}}.
%\]
%We write the total reflection coefficient from the coupler and the
%transmission lines $r$ as follows
% \[
% r=\frac{\frac{\left(Z_{1}-Z_{0}\right)\omega C+j\left(\omega^{2}LC-1\right)}{\left(Z_{1}+Z_{0}\right)\omega C+j\left(\omega^{2}LC-1\right)}\left(e^{2ikl_{2}}r_{2}+1\right)+e^{2ikl_{1}}\left(r_{2}+e^{2ikl_{2}}\right)}{1+e^{2ikl_{2}}r_{2}+e^{2ikl_{1}}\left(r_{2}+e^{2ikl_{2}}\right)\frac{\left(Z_{1}-Z_{0}\right)\omega C+j\left(\omega^{2}LC-1\right)}{\left(Z_{1}+Z_{0}\right)\omega C+j\left(\omega^{2}LC-1\right)}},
% \]
% We first
% impose $g=0,$ from which we express $r_{2},$ the reflection coefficient
% at the $Z_{1}-Z_{2}$ interface, see Supplementary Material for details.
% Since $r_{2}$ is real, we substitute the expression for $\mathrm{Re}\left(r_{2}\right)$
% in $\frac{dg}{d\omega}=0$ and impose $\mathrm{Im}\left(r_{2}\right)=0,$
% from which we express $Z_{1}$ or $C,$ that we also substitute in
% $\frac{dg}{d\omega}=0$. Note that since we only assumed that a transmission
% line has a real impedance, which is almost always the case, this approach
% may apply both to a load before and after the transmission lines or
% both.

Interestingly, when expressing a variable from $\mathrm{Im}\left(r_{2}\right)=0$ (we chose to express $C$) it has two solutions with the form $a\pm \sqrt{b},$ which may indicate that there is an EP4 when $b$ vanishes. We select one of the solutions, fix $\omega_{d} = 1/\sqrt{LC}=2\pi \cdot 5 \, \mathrm{GHz},$ and proceed to get an EP equation for our setup
$\frac{dg}{d\omega}\left(\omega,l_{1,}l_{2},Z_{1},\right)=0.$
%COULD we use this to design a cavity with an EP3?
% Would that be Z_l (letter l) not Z_1 (number one)? If it is true, I don't get it since you already simplified the relation with r_3 = -1, making Z_l fixed. Also, I would prefer having it in the same order as the parameter you showed in Eq. (4) -- $\frac{dg}{d\omega}\left(\omega, Z_{l}, l_{1}, l_{2}, Z_{1}\right)=0.$
% The above process is done by fixing \omega_d and choosing C only, so the dependency of Z_k is gone, am I right?
% AF: it's Z1 (one), changed the order, yes you are right
%Then would the dependency of C be added in \frac{dg}{d\omega} too?
% Also, the way you can find an EP is by varying C to solve \frac{dg}{d\omega} = 0? It is not clear how you can solve \frac{dg}{d\omega}, so I can't revise the above sentences yet.
%We then can express $Z_1$ and 

To solve the EP equation semi-analytically, we set $Z_{0}=50 \, \Omega$ and choose $Z_{1}=100 \Omega$.
We then obtain
$\mathrm{Re}\left(\omega_{\mathrm{EP}}\right) = 2\pi \cdot 4.727 \,\mathrm{GHz},\,\,\,\mathrm{Im}\left(\omega_{\mathrm{EP}}\right)=2\pi \cdot 0.834 \, \mathrm{GHz}, $   $l_{1}=0.0215\mathrm{m},\,\,\,l_{2}=0.0276\mathrm{m},\,\,\,C=0.3037\,\mathrm{pF},\,\,\,Z_{2}=60.27\,\Omega.$
To verify the EP calculation, we plotted $\left|r\right|$ as a function
of $\mathrm{Re}\left(\omega\right)$ and $\mathrm{Im}\left(\omega\right)$ and present it in Fig. 3 (a). It can be seen that there is a clear quadratic dependency in two dimensions, for such a dependency on the real$-\omega$ axis see Ref \cite{sweeney2019perfectly}. In Fig. 3 (b) we plot the coalescence of absorbing eigenfrequencies at the EP, which demonstrates the remarkable accuracy of our EP calculation.  

\begin{figure} 

 \subfloat[]{\includegraphics[width=8cm]{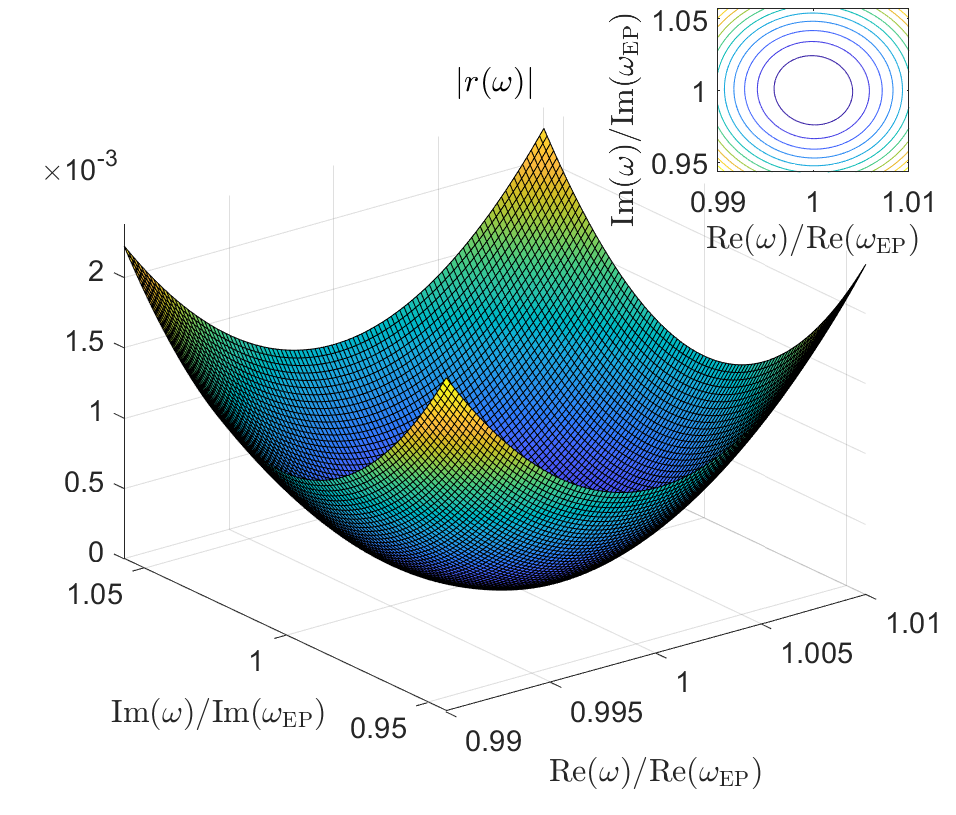}
 }

 \subfloat[]{\includegraphics[width=8cm]{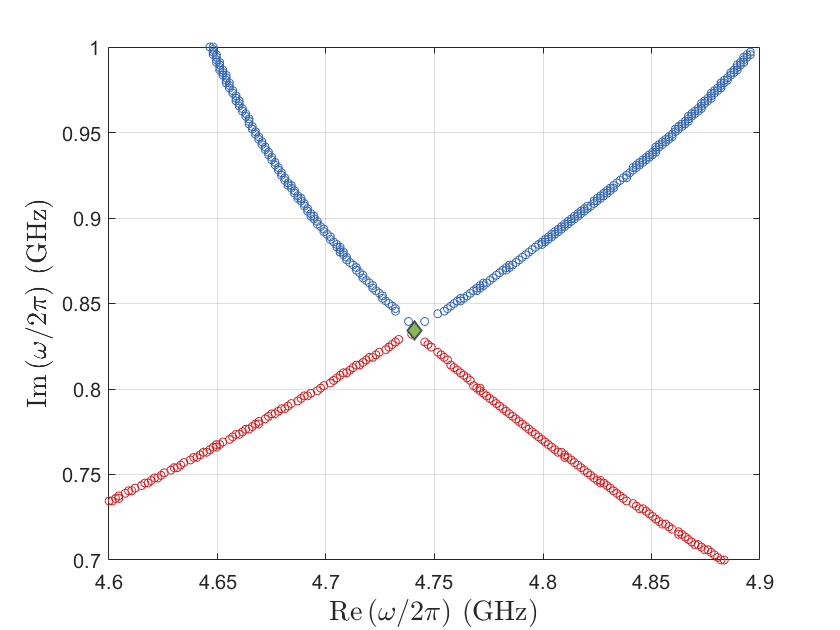}
 }

\caption{ (a) $\left|r\right|$ as a function of
$\mathrm{Re}\left(\omega\right)$ and $\mathrm{Im}\left(\omega\right)$ close to the EP in a surface and contour (inset) plots. The scaling is quadratic in 2D as expected since at the EP $|r|\propto (\omega-\omega_n)^2$ and the
derivative with respect to a complex variable is equal to the derivatives
from all directions, which verifies the calculation of the virtual CPA EP. (b) Coalescence of the eigenfrequencies that satisfy the equation $r=0$ at the EP when varying $l_1/l_2$, where the diamond denotes the EP.}
\end{figure}
%In the Supplementary Material (SM) we calculate additional EPs with
%the standard numerical approach of tracking $\omega_{n}$ when varying
%$l_{1},l_{2}.$ 

%\subsection*{Scattered Fields and Energies}

\begin{figure}
\subfloat[]{

\includegraphics[width=8cm]{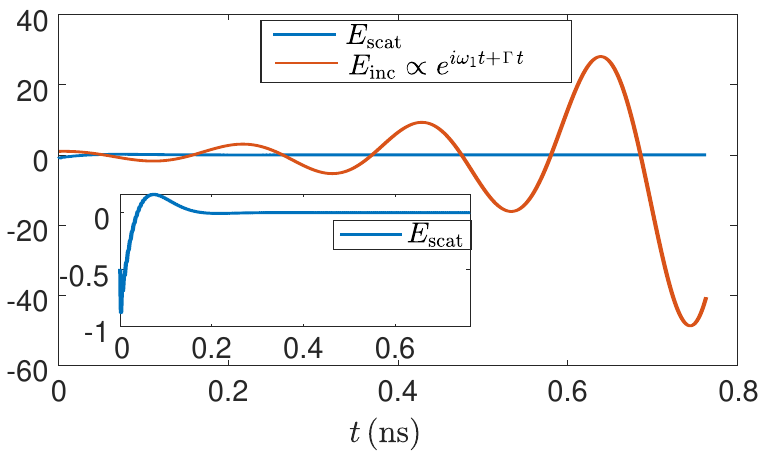}

}

\subfloat[]{\includegraphics[width=8cm]{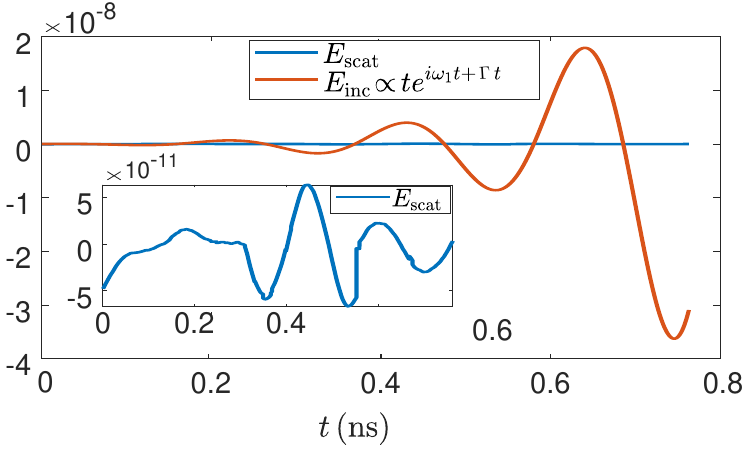}

}

\subfloat[]{\includegraphics[width=8cm]{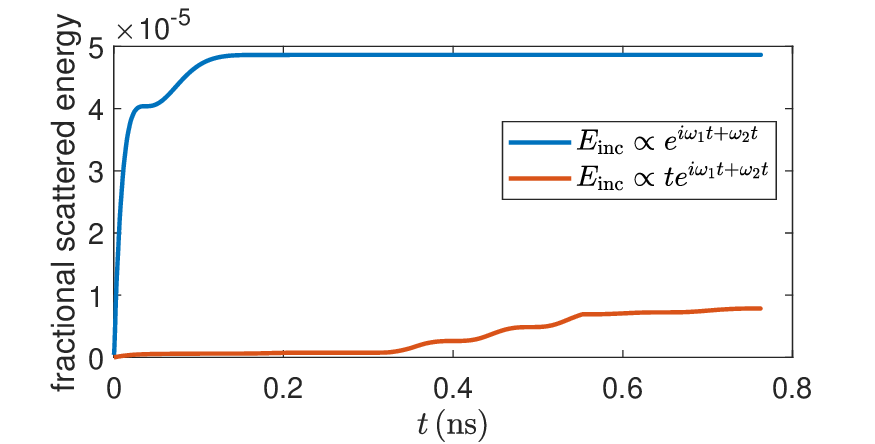}

}

\caption{The incoming and scattered fields as functions of time for the inputs
(a) $e^{i\omega t+\Gamma t}$  
and (b) $te^{i\omega t+\Gamma t}.$
(c) Fractional scattered energy for both incoming fields as functions
of time. Note that for the EP parameters, the roundtrip is less than
an oscillation period, resulting in rapid equilibration.}
\end{figure}

\begin{figure}

\subfloat[]{

\includegraphics[width=8cm]{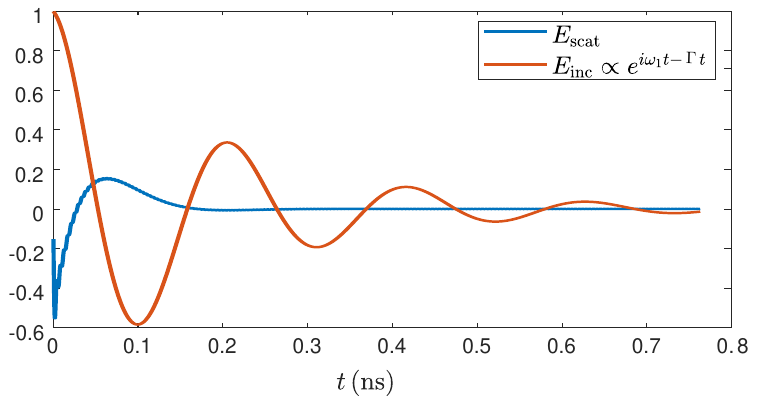}

}

%\subfloat[]{\includegraphics[scale=0.5]{emitted_t_inc_scat_EP}

%}

\subfloat[]{\includegraphics[width=8cm]{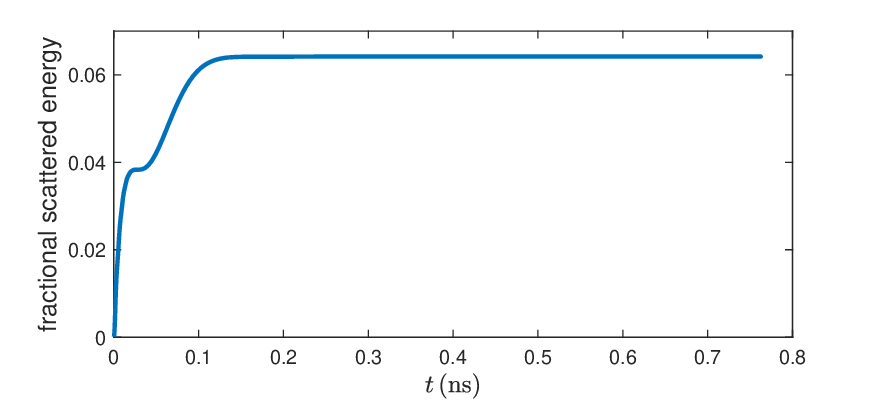}

}

\caption{(a) The incoming and scattered fields as functions of time for the inputs
$e^{i\omega t-\Gamma t}$, which shows 
%and $te^{i\omega t-\Gamma t}$ (b). 
that the
naturally emitted wave $e^{i\omega t-\Gamma t}$ is well captured
by the CPA EP cavity.
%whereas $te^{i\omega t-\Gamma t}$ results in relatively
%large scattering since there is capturing of two orders and one order
%in the expansion, respectively. 
(b) Fractional scattered energy as
a function of time. Importantly, the efficiency for this passive quantum
state transfer is $>93\%.$ Note that there are only two-three roundtrips, and therefore no significant
scattering is expected for the input $e^{i\omega t-\Gamma t}$ during
the second half since there is a delay between the input and output and the linear
 approximation holds in the first half.} 
% (the linear approximation should still hold, note
% that there is a delay between the input and output so that the linear
% approximation needs to hold in the first half).}

\end{figure}

%Let us consider the three different excitation scenarios to quantify the capturing efficiency at the virtual CPA EP: $e^{i\omega t+\Gamma t},te^{i\omega t+\Gamma t}$ and $e^{i\omega t-\Gamma t}$.
To quantify the capturing efficiencies of the exponentially increasing and naturally emitted inputs  at the virtual CPA EP, we calculated the scattered fields.
This calculation was performed using numerical inverse Fourier
Transform by definition since our lossless system has real refractive indices and there are no numerical divergences (topics that were addressed in Ref. \cite{farhi2022excitation}). 
%It seemed "by definition" was just a mistake from earlier version?
In Fig. 4 we present the scattered fields and fractional scattered
energies for the exponentially growing inputs $e^{i\omega t+\Gamma t}$ and $te^{i\omega t+\Gamma t}.$
The fractional scattered energies are $5\cdot10^{-5}$
and $10^{-5}$, respectively, which outperform the efficiencies reported in Ref. \cite{wennerCatchingTimeReversedMicrowave2014}, and exceed the fidelities
required for good logic gates and measurements.

In Fig. 5 we present the scattered fields and fractional scattered
energies for the naturally emitted wave $e^{i\omega t-\Gamma t}.$
Importantly, while the previously
reported fractional scattering for such wave is 39\% \cite{wennerCatchingTimeReversedMicrowave2014},
in our case, it is less than 7\%, even though our Q-factor is larger by a factor of 2.61, which is expected to reduce the performance (e.g., due to a large $r_1$). This fractional scattering can be further decreased by reducing
the value of $r_{1}.$ To confirm this we calculated another EP with 1.63 times lower Q factor and obtained 4.5\% fractional energy, see Supplementary Material (SM) for details. We can thus extrapolate that using our approach for the same Q factor a first-order CPA EP will perform approximately an order of magnitude better. In the SM we also plot the scattered field for the drive $te^{i\omega t-\Gamma t}$, which has a relatively large scattered field since the system captures only one temporal order of
the input. As noted, this process does not require modulation of the coupler, which holds potential to enable applications at optical frequencies. Similarly, photon detection of waveforms emitted in many processes such as spontaneous emission can be significantly improved.

In summary, we first showed that an absorbing exceptional point captures additional temporal orders of any incoming waveform. We presented a general approach to tune the receiving cavity to an exceptional point, which considerably simplifies a highly complex problem and achieves a very high calculation accuracy. We then demonstrated that this system is able to perform significantly better compared with the existing systems in terms of catching efficiency, surpassing the required fidelities for certain applications. Furthermore, we showed that a virtual CPA EP port can efficiently catch naturally emitted photons, potentially opening avenues for quantum and photonic computation at optical frequencies and efficient photon detection.  It is important to note that large $\Gamma\mathrm{s}$ of the receiving port improve the performance, even it increases the distance to complex $\omega$ of the incoming wave and requires a faster switch off. It is also worth mentioning that our approach is  valid for lossy systems, which may be important e.g., for room temperature applications. We expect that our results will apply to other fields of wave physics such as acoustics and matter waves \cite{chu2017quantum,manenti2017circuit,mullers2018coherent,soley2023experimentally}.
% Do you have citations for matter waves?
% AF: added two citations
\\
\\
% Action items:
% \begin{itemize}
% \item Done: A general approach to calculate EPs for transmission lines + coupler + qubit that does not depend on the details of the lumped elements.  
% \item Done: Ask Seunghwi/Andrea about the analogue of a lumped element,  $Z_k[\omega]$ at infrared optical frequencies, photonic circuits.
% \item Done: Fig. 1 for passive state transfer
% \item Done: Wei: Fig. 2 for the general coupler, transmission lines and qubit setup. Literature survey on quantum computing and intro. paragraph.
% \item  Done: Fig. on coalescence of eigevalues from Seunghwi.
% \item Done: Make Fig. 3 (a) two dimensional.
% \item Done: making 3 c 4 c more compact, see if 4 b is important (move to SM).
% \end{itemize}
%\section{Acknowledgements}
We acknowledge the fruitful discussions with V. Joshi,  M. Devoret, and R. Cortinas. This work was partially supported by a grant from the Simons
Foundation.
\bibliographystyle{unsrt}

\begin{thebibliography}{10}

\bibitem{kimbleQuantumInternet2008}
H.~J. Kimble.
\newblock The quantum internet.
\newblock {\em Nature}, 453(7198):1023--1030, June 2008.

\bibitem{northupQuantumInformationTransfer2014}
T.~E. Northup and R.~Blatt.
\newblock Quantum information transfer using photons.
\newblock {\em Nature Photonics}, 8(5):356--363, May 2014.

\bibitem{monroeLargescaleModularQuantumcomputer2014}
C.~Monroe, R.~Raussendorf, A.~Ruthven, K.~R. Brown, P.~Maunz, L.-M. Duan, and
  J.~Kim.
\newblock Large-scale modular quantum-computer architecture with atomic memory
  and photonic interconnects.
\newblock {\em Physical Review A}, 89(2):022317, February 2014.

\bibitem{jiangDistributedQuantumComputation2007}
Liang Jiang, Jacob~M. Taylor, Anders~S. S{\o}rensen, and Mikhail~D. Lukin.
\newblock Distributed quantum computation based on small quantum registers.
\newblock {\em Physical Review A}, 76(6):062323, December 2007.

\bibitem{duanLongdistanceQuantumCommunication2001}
L.-M. Duan, M.~D. Lukin, J.~I. Cirac, and P.~Zoller.
\newblock Long-distance quantum communication with atomic ensembles and linear
  optics.
\newblock {\em Nature}, 414(6862):413--418, November 2001.

\bibitem{childressFaultTolerantQuantumCommunication2006}
L.~Childress, J.~M. Taylor, A.~S. S{\o}rensen, and M.~D. Lukin.
\newblock Fault-{{Tolerant Quantum Communication Based}} on {{Solid-State
  Photon Emitters}}.
\newblock {\em Physical Review Letters}, 96(7):070504, February 2006.

\bibitem{ciracQuantumStateTransfer1997}
J.~I. Cirac, P.~Zoller, H.~J. Kimble, and H.~Mabuchi.
\newblock Quantum {{State Transfer}} and {{Entanglement Distribution}} among
  {{Distant Nodes}} in a {{Quantum Network}}.
\newblock {\em Physical Review Letters}, 78(16):3221--3224, April 1997.

\bibitem{chouMeasurementinducedEntanglementExcitation2005}
C.~W. Chou, H.~{de Riedmatten}, D.~Felinto, S.~V. Polyakov, S.~J. {van Enk},
  and H.~J. Kimble.
\newblock Measurement-induced entanglement for excitation stored in remote
  atomic ensembles.
\newblock {\em Nature}, 438(7069):828--832, December 2005.

\bibitem{moehringEntanglementSingleatomQuantum2007}
D.~L. Moehring, P.~Maunz, S.~Olmschenk, K.~C. Younge, D.~N. Matsukevich, L.-M.
  Duan, and C.~Monroe.
\newblock Entanglement of single-atom quantum bits at a distance.
\newblock {\em Nature}, 449(7158):68--71, September 2007.

\bibitem{hofmannHeraldedEntanglementWidely2012}
Julian Hofmann, Michael Krug, Norbert Ortegel, Lea G{\'e}rard, Markus Weber,
  Wenjamin Rosenfeld, and Harald Weinfurter.
\newblock Heralded {{Entanglement Between Widely Separated Atoms}}.
\newblock {\em Science}, 337(6090):72--75, July 2012.

\bibitem{bernienHeraldedEntanglementSolidstate2013}
H.~Bernien, B.~Hensen, W.~Pfaff, G.~Koolstra, M.~S. Blok, L.~Robledo, T.~H.
  Taminiau, M.~Markham, D.~J. Twitchen, L.~Childress, and R.~Hanson.
\newblock Heralded entanglement between solid-state qubits separated by three
  metres.
\newblock {\em Nature}, 497(7447):86--90, May 2013.

\bibitem{olmschenkQuantumTeleportationDistant2009}
S.~Olmschenk, D.~N. Matsukevich, P.~Maunz, D.~Hayes, L.-M. Duan, and C.~Monroe.
\newblock Quantum {{Teleportation Between Distant Matter Qubits}}.
\newblock {\em Science}, 323(5913):486--489, January 2009.

\bibitem{ritterElementaryQuantumNetwork2012}
Stephan Ritter, Christian N{\"o}lleke, Carolin Hahn, Andreas Reiserer, Andreas
  Neuzner, Manuel Uphoff, Martin M{\"u}cke, Eden Figueroa, Joerg Bochmann, and
  Gerhard Rempe.
\newblock An elementary quantum network of single atoms in optical cavities.
\newblock {\em Nature}, 484(7393):195--200, April 2012.

\bibitem{kurpiersDeterministicQuantumState2018}
P.~Kurpiers, P.~Magnard, T.~Walter, B.~Royer, M.~Pechal, J.~Heinsoo,
  Y.~Salath{\'e}, A.~Akin, S.~Storz, J.-C. Besse, S.~Gasparinetti, A.~Blais,
  and A.~Wallraff.
\newblock Deterministic quantum state transfer and remote entanglement using
  microwave photons.
\newblock {\em Nature}, 558(7709):264--267, June 2018.

\bibitem{axlineOndemandQuantumState2018}
Christopher~J. Axline, Luke~D. Burkhart, Wolfgang Pfaff, Mengzhen Zhang, Kevin
  Chou, Philippe {Campagne-Ibarcq}, Philip Reinhold, Luigi Frunzio, S.~M.
  Girvin, Liang Jiang, M.~H. Devoret, and R.~J. Schoelkopf.
\newblock On-demand quantum state transfer and entanglement between remote
  microwave cavity memories.
\newblock {\em Nature Physics}, 14(7):705--710, July 2018.

\bibitem{campagne-ibarcqDeterministicRemoteEntanglement2018}
P.~{Campagne-Ibarcq}, E.~{Zalys-Geller}, A.~Narla, S.~Shankar, P.~Reinhold,
  L.~Burkhart, C.~Axline, W.~Pfaff, L.~Frunzio, R.~J. Schoelkopf, and M.~H.
  Devoret.
\newblock Deterministic {{Remote Entanglement}} of {{Superconducting Circuits}}
  through {{Microwave Two-Photon Transitions}}.
\newblock {\em Physical Review Letters}, 120(20):200501, May 2018.

\bibitem{flurinSuperconductingQuantumNode2015}
E.~Flurin, N.~Roch, J.~D. Pillet, F.~Mallet, and B.~Huard.
\newblock Superconducting {{Quantum Node}} for {{Entanglement}} and {{Storage}}
  of {{Microwave Radiation}}.
\newblock {\em Physical Review Letters}, 114(9):090503, March 2015.

\bibitem{yinCatchReleaseMicrowave2013}
Yi~Yin, Yu~Chen, Daniel Sank, P.~J.~J. O'Malley, T.~C. White, R.~Barends,
  J.~Kelly, Erik Lucero, Matteo Mariantoni, A.~Megrant, C.~Neill,
  A.~Vainsencher, J.~Wenner, Alexander~N. Korotkov, A.~N. Cleland, and John~M.
  Martinis.
\newblock Catch and {{Release}} of {{Microwave Photon States}}.
\newblock {\em Physical Review Letters}, 110(10):107001, March 2013.

\bibitem{reisererCavitybasedQuantumNetworks2015}
Andreas Reiserer and Gerhard Rempe.
\newblock Cavity-based quantum networks with single atoms and optical photons.
\newblock {\em Reviews of Modern Physics}, 87(4):1379--1418, December 2015.

\bibitem{goy1983observation}
Ph~Goy, JM~Raimond, M~Gross, and S~Haroche.
\newblock Observation of cavity-enhanced single-atom spontaneous emission.
\newblock {\em Physical review letters}, 50(24):1903, 1983.

\bibitem{albertinaleDetectingSpinsTheir2021}
Emanuele Albertinale, L{\'e}o Balembois, Eric Billaud, Vishal Ranjan, Daniel
  Flanigan, Thomas Schenkel, Daniel Est{\`e}ve, Denis Vion, Patrice Bertet, and
  Emmanuel Flurin.
\newblock Detecting spins by their fluorescence with a microwave photon
  counter.
\newblock {\em Nature}, 600(7889):434--438, December 2021.

\bibitem{wangSingleelectronSpinResonance2023}
Z.~Wang, L.~Balembois, M.~Ran{\v c}i{\'c}, E.~Billaud, M.~Le~Dantec,
  A.~Ferrier, P.~Goldner, S.~Bertaina, T.~Chaneli{\`e}re, D.~Esteve, D.~Vion,
  P.~Bertet, and E.~Flurin.
\newblock Single-electron spin resonance detection by microwave photon
  counting.
\newblock {\em Nature}, 619(7969):276--281, July 2023.

\bibitem{graham2022multi}
TM~Graham, Y~Song, J~Scott, C~Poole, L~Phuttitarn, K~Jooya, P~Eichler, X~Jiang,
  A~Marra, B~Grinkemeyer, et~al.
\newblock Multi-qubit entanglement and algorithms on a neutral-atom quantum
  computer.
\newblock {\em Nature}, 604(7906):457--462, 2022.

\bibitem{browaeys2020many}
Antoine Browaeys and Thierry Lahaye.
\newblock Many-body physics with individually controlled rydberg atoms.
\newblock {\em Nature Physics}, 16(2):132--142, 2020.

\bibitem{hadfield2009single}
Robert~H Hadfield.
\newblock Single-photon detectors for optical quantum information applications.
\newblock {\em Nature photonics}, 3(12):696--705, 2009.

\bibitem{wennerCatchingTimeReversedMicrowave2014}
J.~Wenner, Yi~Yin, Yu~Chen, R.~Barends, B.~Chiaro, E.~Jeffrey, J.~Kelly,
  A.~Megrant, J.~Y. Mutus, C.~Neill, P.~J.~J. O'Malley, P.~Roushan, D.~Sank,
  A.~Vainsencher, T.~C. White, Alexander~N. Korotkov, A.~N. Cleland, and
  John~M. Martinis.
\newblock Catching {{Time-Reversed Microwave Coherent State Photons}} with
  99.4\% {{Absorption Efficiency}}.
\newblock {\em Physical Review Letters}, 112(21):210501, May 2014.

\bibitem{palomakiCoherentStateTransfer2013}
T.~A. Palomaki, J.~W. Harlow, J.~D. Teufel, R.~W. Simmonds, and K.~W. Lehnert.
\newblock Coherent state transfer between itinerant microwave fields and a
  mechanical oscillator.
\newblock {\em Nature}, 495(7440):210--214, March 2013.

\bibitem{linDeterministicLoadingMicrowaves2022}
Wei-Ju Lin, Yong Lu, Ping~Yi Wen, Yu-Ting Cheng, Ching-Ping Lee, Kuan~Ting Lin,
  Kuan~Hsun Chiang, Ming~Che Hsieh, Ching-Yeh Chen, Chin-Hsun Chien, Jia~Jhan
  Lin, Jeng-Chung Chen, Yen~Hsiang Lin, Chih-Sung Chuu, Franco Nori, Anton
  Frisk~Kockum, Guin~Dar Lin, Per Delsing, and Io-Chun Hoi.
\newblock Deterministic {{Loading}} of {{Microwaves}} onto an {{Artificial Atom
  Using}} a {{Time-Reversed Waveform}}.
\newblock {\em Nano Letters}, 22(20):8137--8142, October 2022.

\bibitem{baderEfficientCouplingOptical2013}
M.~Bader, S.~Heugel, A.~L. Chekhov, M.~Sondermann, and G.~Leuchs.
\newblock Efficient coupling to an optical resonator by exploiting
  time-reversal symmetry.
\newblock {\em New Journal of Physics}, 15(12):123008, December 2013.

\bibitem{liuEfficientlyLoadingSingle2014}
Chang Liu, Yuan Sun, Luwei Zhao, Shanchao Zhang, M.~M.~T. Loy, and Shengwang
  Du.
\newblock Efficiently {{Loading}} a {{Single Photon}} into a {{Single-Sided
  Fabry-Perot Cavity}}.
\newblock {\em Physical Review Letters}, 113(13):133601, September 2014.

\bibitem{baranov2017coherent}
Denis~G Baranov, Alex Krasnok, and Andrea Alu.
\newblock Coherent virtual absorption based on complex zero excitation for
  ideal light capturing.
\newblock {\em Optica}, 4(12):1457--1461, 2017.

\bibitem{chong2010coherent}
YD~Chong, Li~Ge, Hui Cao, and A~Douglas Stone.
\newblock Coherent perfect absorbers: time-reversed lasers.
\newblock {\em Physical review letters}, 105(5):053901, 2010.

\bibitem{noh2012perfect}
Heeso Noh, Yidong Chong, A~Douglas Stone, and Hui Cao.
\newblock Perfect coupling of light to surface plasmons by coherent absorption.
\newblock {\em Physical review letters}, 108(18):186805, 2012.

\bibitem{khanahmadi2023multimode}
Maryam Khanahmadi, Mads~Middelhede Lund, Klaus M{\o}lmer, and G{\"o}ran
  Johansson.
\newblock The multimode character of quantum states released from a
  superconducting cavity.
\newblock {\em arXiv preprint arXiv:2306.12127}, 2023.

\bibitem{korotkovFlyingMicrowaveQubits2011}
Alexander~N. Korotkov.
\newblock Flying microwave qubits with nearly perfect transfer efficiency.
\newblock {\em Physical Review B}, 84(1):014510, July 2011.

\bibitem{seteRobustQuantumState2015}
Eyob~A. Sete, Eric Mlinar, and Alexander~N. Korotkov.
\newblock Robust quantum state transfer using tunable couplers.
\newblock {\em Physical Review B}, 91(14):144509, April 2015.

\bibitem{srinivasanTimereversalSymmetrizationSpontaneous2014}
Srikanth~J. Srinivasan, Neereja~M. Sundaresan, Darius Sadri, Yanbing Liu,
  Jay~M. Gambetta, Terri Yu, S.~M. Girvin, and Andrew~A. Houck.
\newblock Time-reversal symmetrization of spontaneous emission for quantum
  state transfer.
\newblock {\em Physical Review A}, 89(3):033857, March 2014.

\bibitem{pierreStorageOndemandRelease2014}
Mathieu Pierre, Ida-Maria Svensson, Sankar Raman~Sathyamoorthy, G{\"o}ran
  Johansson, and Per Delsing.
\newblock Storage and on-demand release of microwaves using superconducting
  resonators with tunable coupling.
\newblock {\em Applied Physics Letters}, 104(23):232604, June 2014.

\bibitem{pfaffControlledReleaseMultiphoton2017}
Wolfgang Pfaff, Christopher~J. Axline, Luke~D. Burkhart, Uri Vool, Philip
  Reinhold, Luigi Frunzio, Liang Jiang, Michel~H. Devoret, and Robert~J.
  Schoelkopf.
\newblock Controlled release of multiphoton quantum states from a microwave
  cavity memory.
\newblock {\em Nature Physics}, 13(9):882--887, September 2017.

\bibitem{pechalMicrowaveControlledGenerationShaped2014}
M.~Pechal, L.~Huthmacher, C.~Eichler, S.~Zeytino{\u g}lu, A.~A. Abdumalikov,
  S.~Berger, A.~Wallraff, and S.~Filipp.
\newblock Microwave-{{Controlled Generation}} of {{Shaped Single Photons}} in
  {{Circuit Quantum Electrodynamics}}.
\newblock {\em Physical Review X}, 4(4):041010, October 2014.

\bibitem{mazor2021unitary}
Yarden Mazor, Michele Cotrufo, and Andrea Alu.
\newblock Unitary excitation transfer between coupled cavities using temporal
  switching.
\newblock {\em Physical Review Letters}, 127(1):013902, 2021.

\bibitem{houckGeneratingSingleMicrowave2007}
A.~A. Houck, D.~I. Schuster, J.~M. Gambetta, J.~A. Schreier, B.~R. Johnson,
  J.~M. Chow, L.~Frunzio, J.~Majer, M.~H. Devoret, S.~M. Girvin, and R.~J.
  Schoelkopf.
\newblock Generating single microwave photons in a circuit.
\newblock {\em Nature}, 449(7160):328--331, September 2007.

\bibitem{kindelGenerationEfficientMeasurement2016}
William~F. Kindel, M.~D. Schroer, and K.~W. Lehnert.
\newblock Generation and efficient measurement of single photons from
  fixed-frequency superconducting qubits.
\newblock {\em Physical Review A}, 93(3):033817, March 2016.

\bibitem{heugel2010analogy}
Simon Heugel, Alessandro~S Villar, Markus Sondermann, Ulf Peschel, and Gerd
  Leuchs.
\newblock On the analogy between a single atom and an optical resonator.
\newblock {\em Laser Physics}, 20:100--106, 2010.

\bibitem{bender1998real}
Carl~M Bender and Stefan Boettcher.
\newblock Real spectra in non-hermitian hamiltonians having p t symmetry.
\newblock {\em Physical Review Letters}, 80(24):5243, 1998.

\bibitem{makris2008beam}
Konstantinos~G Makris, R~El-Ganainy, DN~Christodoulides, and Ziad~H Musslimani.
\newblock Beam dynamics in pt-symmetric optical lattices.
\newblock {\em Physical Review Letters}, 100(10):103904, 2008.

\bibitem{moiseyev2011non}
Nimrod Moiseyev.
\newblock {\em Non-Hermitian quantum mechanics}.
\newblock Cambridge University Press, 2011.

\bibitem{miri2019exceptional}
Mohammad-Ali Miri and Andrea Alu.
\newblock Exceptional points in optics and photonics.
\newblock {\em Science}, 363(6422):eaar7709, 2019.

\bibitem{ruter2010observation}
Christian~E R{\"u}ter, Konstantinos~G Makris, Ramy El-Ganainy, Demetrios~N
  Christodoulides, Mordechai Segev, and Detlef Kip.
\newblock Observation of parity--time symmetry in optics.
\newblock {\em Nature physics}, 6(3):192--195, 2010.

\bibitem{sweeney2019perfectly}
William~R Sweeney, Chia~Wei Hsu, Stefan Rotter, and A~Douglas Stone.
\newblock Perfectly absorbing exceptional points and chiral absorbers.
\newblock {\em Physical review letters}, 122(9):093901, 2019.

\bibitem{wang2021coherent}
Changqing Wang, William~R Sweeney, A~Douglas Stone, and Lan Yang.
\newblock Coherent perfect absorption at an exceptional point.
\newblock {\em Science}, 373(6560):1261--1265, 2021.

\bibitem{farhi2022excitation}
Asaf Farhi, Ahmed Mekawy, Andrea Alu, and Douglas Stone.
\newblock Excitation of absorbing exceptional points in the time domain.
\newblock {\em Physical Review A}, 106(3):L031503, 2022.

\bibitem{mekawy2023observation}
Ahmed Mekawy, Asaf Farhi, Douglas Stone, and Andrea Alu.
\newblock Observation of absorbing exceptional points in the time domain.
\newblock To be submitted.

\bibitem{pozar2011microwave}
David~M Pozar.
\newblock {\em Microwave engineering}.
\newblock John wiley \& sons, 2011.

\bibitem{engheta2005circuit}
Nader Engheta, Alessandro Salandrino, and Andrea Alu.
\newblock Circuit elements at optical frequencies: nanoinductors,
  nanocapacitors, and nanoresistors.
\newblock {\em Physical Review Letters}, 95(9):095504, 2005.

\bibitem{engheta2007circuits}
Nader Engheta.
\newblock Circuits with light at nanoscales: optical nanocircuits inspired by
  metamaterials.
\newblock {\em science}, 317(5845):1698--1702, 2007.

\bibitem{alu2008tuning}
Andrea Alu and Nader Engheta.
\newblock Tuning the scattering response of optical nanoantennas with
  nanocircuit loads.
\newblock {\em Nature photonics}, 2(5):307--310, 2008.

\bibitem{schnell2009controlling}
Martin Schnell, Aitzol Garc{\'\i}a-Etxarri, AJ~Huber, K~Crozier, Javier
  Aizpurua, and Rainer Hillenbrand.
\newblock Controlling the near-field oscillations of loaded plasmonic
  nanoantennas.
\newblock {\em Nature Photonics}, 3(5):287--291, 2009.

\bibitem{yao2013broad}
Yu~Yao, Mikhail~A Kats, Patrice Genevet, Nanfang Yu, Yi~Song, Jing Kong, and
  Federico Capasso.
\newblock Broad electrical tuning of graphene-loaded plasmonic antennas.
\newblock {\em Nano letters}, 13(3):1257--1264, 2013.

\bibitem{liu2013individual}
Na~Liu, Fangfang Wen, Yang Zhao, Yumin Wang, Peter Nordlander, Naomi~J Halas,
  and Andrea Al{\`u}.
\newblock Individual nanoantennas loaded with three-dimensional optical
  nanocircuits.
\newblock {\em Nano letters}, 13(1):142--147, 2013.

\bibitem{aouani2014third}
Heykel Aouani, Mohsen Rahmani, Miguel Navarro-C{\'\i}a, and Stefan~A Maier.
\newblock Third-harmonic-upconversion enhancement from a single semiconductor
  nanoparticle coupled to a plasmonic antenna.
\newblock {\em Nature nanotechnology}, 9(4):290--294, 2014.

\bibitem{venkataraman2011few}
Vivek Venkataraman, Kasturi Saha, Pablo Londero, and Alexander~L Gaeta.
\newblock Few-photon all-optical modulation in a photonic band-gap fiber.
\newblock {\em Physical review letters}, 107(19):193902, 2011.

\bibitem{ren2011nanostructured}
Mengxin Ren, Baohua Jia, Jun-Yu Ou, Eric Plum, Jianfa Zhang, Kevin~F MacDonald,
  Andrey~E Nikolaenko, Jingjun Xu, Min Gu, and Nikolay~I Zheludev.
\newblock Nanostructured plasmonic medium for terahertz bandwidth all-optical
  switching.
\newblock {\em Advanced Materials}, 23(46):5540--5544, 2011.

\bibitem{chu2017quantum}
Yiwen Chu, Prashanta Kharel, William~H Renninger, Luke~D Burkhart, Luigi
  Frunzio, Peter~T Rakich, and Robert~J Schoelkopf.
\newblock Quantum acoustics with superconducting qubits.
\newblock {\em Science}, 358(6360):199--202, 2017.

\bibitem{manenti2017circuit}
Riccardo Manenti, Anton~F Kockum, Andrew Patterson, Tanja Behrle, Joseph
  Rahamim, Giovanna Tancredi, Franco Nori, and Peter~J Leek.
\newblock Circuit quantum acoustodynamics with surface acoustic waves.
\newblock {\em Nature communications}, 8(1):975, 2017.

\bibitem{mullers2018coherent}
Andreas M{\"u}llers, Bodhaditya Santra, Christian Baals, Jian Jiang, Jens
  Benary, Ralf Labouvie, Dmitry~A Zezyulin, Vladimir~V Konotop, and Herwig Ott.
\newblock Coherent perfect absorption of nonlinear matter waves.
\newblock {\em Science Advances}, 4(8):eaat6539, 2018.

\bibitem{soley2023experimentally}
Micheline~B Soley, Carl~M Bender, and A~Douglas Stone.
\newblock Experimentally realizable p t phase transitions in reflectionless
  quantum scattering.
\newblock {\em Physical Review Letters}, 130(25):250404, 2023.

\end{thebibliography}

\section{Supplementary Material}

We write explicitly $r_2$ for the case of negligible coupling to the transmon during the drive \cite{wennerCatchingTimeReversedMicrowave2014} and the coupler implementation in Fig. 2.
%We validate our calculation for the EP2

%\[
%e^{\Gamma_{2}t}\left(1-\left(\Gamma+\Gamma_{2}\right)t\right)\approx\left(1+\Gamma_{2}t\right)\left(1-\left(\Gamma+\Gamma_{2}\right)t\right)\approx1+\Gamma_{2}t-\left(\Gamma+\Gamma_{2}\right)t=1-\Gamma t.
%\]

\begin{widetext}
\[
r_{2}=\frac{-\left(C_{1}(Z_{1}+Z_{0})(\omega_{r}+i\omega_{i})+i\left(-1+\frac{(\omega_{r}+i\omega_{i})^{2}}{\omega_{d}^{2}}\right)\right)e^{\frac{2il_{1}(\omega_{r}+i\omega_{i})}{v}+\frac{2il_{2}(\omega_{r}+i\omega_{i})}{v}}-C_{1}(Z_{1}-Z_{0})(\omega_{r}+i\omega_{i})-i\left(-1+\frac{(\omega_{r}+i\omega_{i})^{2}}{\text{\ensuremath{\omega_{d}}}^{2}}\right)}{e^{\frac{2il_{1}(\omega_{r}+i\omega_{i})}{v}}\left(C_{1}(Z_{1}+Z_{0})(\omega_{r}+i\omega_{i})+i\left(-1+\frac{(\omega_{r}+i\omega_{i}){}^{2}}{\text{\ensuremath{\omega_{d}}}^{2}}\right)\right)+e^{\frac{2il_{2}(\omega_{r}+i\omega_{i})}{v}}\left(C_{1}(Z_{1}-Z_{0})(\omega_{r}+i\omega_{i}))+i\left(-1+\frac{(\omega_{r}+i\omega_{i})^{2}}{\text{\ensuremath{\omega_{d}}}^{2}}\right)\right)}.
\]
\end{widetext}
To quantify the effect of varying the Q-factor on the performance
we calculated another EP with 1.63 times lower Q-factor and the following parameters:
\begin{gather}
    \mathrm{Re}(\omega_\mathrm{EP})= 5.13\cdot 2\pi \,\mathrm{(GHz)},\,\,
    \mathrm{Im}(\omega_\mathrm{EP})=1.47\cdot 2\pi \,\mathrm{(GHz)},\nonumber\\
 Z_1=80 \Omega,\,\,l_1= 0.0195,\,\l_2= 0.0175,\nonumber\\
C =0.867\, (\mathrm{pF}),\,\,
r_2=0.4775,\nonumber
\end{gather}
and obtained 4.5\% fractional energy, which confirms the expected dependency.

In Fig. 6 we present the scattered field for the input $te^{i\omega t-\Gamma t}$, which has a relatively large scattered field since the system captures only one temporal order of the input as expected. 
\begin{figure}

\includegraphics[scale=0.5]{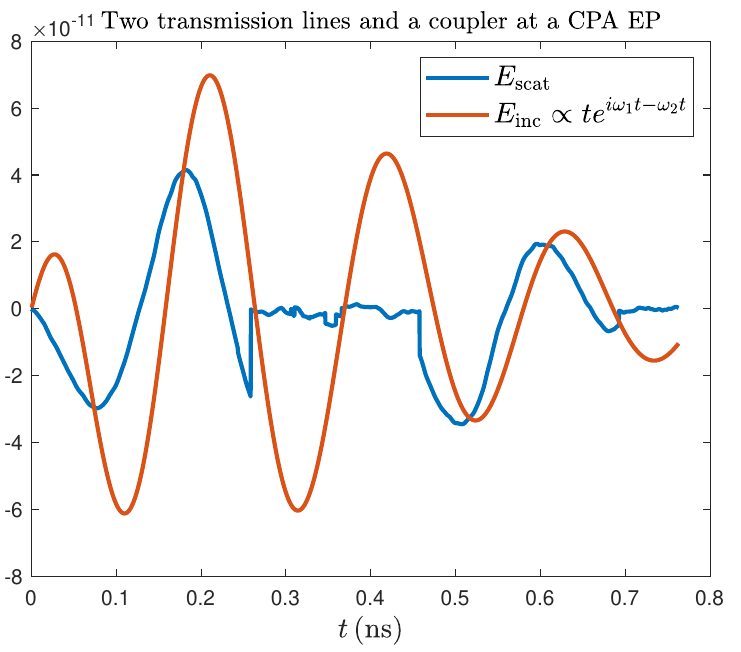}

\caption{Scattering of an incoming wave $te^{i\omega t-\omega_2 t}$ at a virtual CPA EP. It can be seen that since this waveform expansion starts from linear order, only one order is captured, which leads to considerable scattering.}

\end{figure}

% \subsubsection*{Standard approach to calculate EPs}

% Calculating eigenvalues and searching for an EP bahavior and then
% coalesence of eigenvalues

% \begin{figure}
% \includegraphics[scale=0.35]{\string"real w as a function of l1 and l2 zoomed\string".pdf}\includegraphics[scale=0.35]{\string"Im (w) EP 5GHz real freq\string".pdf}\includegraphics[scale=0.35]{\string"EP 2\string".pdf}

% \caption{$\mathrm{Re}\left(\omega\right)$ and $\mathrm{Im}\left(\omega\right)$
% as functions of $l_{1},l_{2}$ for $c=3\cdot10^{8},Z_{0}=50,Z_{1}=70,Z_{2}=210,C=10^{-12},1/\sqrt{LC}=5\cdot10^{9}$}
% \end{figure}

\end{document}